\documentclass[prd,aps]{revtex4-1}

\usepackage{epsfig}
\usepackage{graphicx,color}

\newcommand{\beq}{\begin{equation}}
\newcommand{\eeq}{\end{equation}}
\newcommand{\be}{\begin{eqnarray}}
\newcommand{\ee}{\end{eqnarray}}

\usepackage{ulem}

\begin{document}

\title{
Positronium Decays with Dark $Z$ and Fermionic Dark Matter 
}

\author{ Dong-Won Jung }
\email{dongwonj@korea.ac.kr}

\author{Chaehyun Yu}
\email{chyu@korea.ac.kr}

\affiliation{
Department of Physics,
Korea University, Seoul 02841, Korea
}

\author{ Kang Young Lee }
\email{kylee.phys@gnu.ac.kr}

\affiliation{
Department of Physics Education \& RINS
Gyeongsang National University, Jinju 52828, Korea
}

\date{\today}

\begin{abstract}

We investigate the invisible decay of positronium
to probe the fermionic light dark matter 
mediated by the dark $Z$ boson.
Too tiny is the invisible decay rate of positronium
through weak interaction in the standard model
to be detected in the experiment.
We show that  
it can be enhanced to be observed in the future
if the dark matter is lighter than the electron
in the dark $Z$ model.
We also compute the relic abundance of such light dark matter 
and discuss the Big Bang Nucleosynthesis constraint
with an alternative thermal history scenario.

\end{abstract}

\pacs{ }

\maketitle

\section{Introduction}
Positronium (Ps) is a leptonic atom
which consists of an electron-positron bound state
and the lightest bound state in the standard model (SM).
The wavefunction of the Ps in the leading order 
is obtained by solving Schr\"odinger equation of the hydrogen atom 
with the reduced mass equal to the half of the electron mass.
The lowest states of Ps are 
the spin singlet ($^1S_0$) and the spin triplet states ($^3S_1$),
called para-positronium (p-Ps) and ortho-positronium (o-Ps),
respectively.
The energy spectra and lifetimes of Ps states can be
calculated in QED with high accuracy
since theoretical study of Ps is free from hadronic uncertainty.
Combined with precise measurements,
the study of Ps allows us to test
our understanding of bound state structure of QED~\cite{Ps1,Ps2}.

The dominant decay channel of the p-Ps 
is two-photon decay with the lifetime 
$\tau = 7989.6060(2)^{-1}~\mu{\rm s}$
\cite{pPslifetime},
while that of the o-Ps 
three-photon decay with the lifetime 
$\tau = (7.0380-7.0417)^{-1}~\mu{\rm s}$~\cite{oPslifetime}.
The triplet state o-Ps may decay into neutrino pairs
via weak interaction to leave invisible final state 
in the SM.
However the SM invisible decay rate of the o-Ps is extremely small
such that the branching ratios 
$\sim 6.2 \times 10^{-18}$ (for $\nu_e$) and 
$\sim 9.5 \times 10^{-21}$ (for $\nu_{\mu,\tau}$)
\cite{weakdecays}.
The p-Ps may also decay into neutrino pairs via weak interaction,
but its decay rates are much smaller than those of the o-Ps
because the weak decay rates into neutrinos for the p-Ps
are proportional to the squares of the neutrino masses.
Thus the Ps invisible decays can be a good testing laboratory
of the new physics beyond the SM
in both o-Ps and p-Ps decays.
Actually sizable invisible decays of o-Ps are predicted
in many new physics models,
e.g. millicharged particles, paraphotons etc.
\cite{LDM,NewPhysics,rubbia}.

Experimental searches for invisible decays of Ps
have been performed but observed no signals so far.
The most stringent upper bound on the o-Ps invisible decay
branching ratio is set to be \cite{badertscher}
\be
{\rm Br(o\textrm{-} Ps \to invisible}) < 4.2 \times 10^{-7}
\label{broPs}
\ee
at 90$\%$ C.L..
Recently Vigo et al. \cite{Vigo2020} 
also set the model independent upper limit, 
Br$({\rm Ps} \to {\rm invisible}) < 1.7 \times 10^{-5}$
at 90 \% C.L.
in an alternative experiment. 
Note that decays of o-Ps is more manageable in experiments 
due to its longer lifetime.

One of the most important challenges in particle physics at present
is resolving nature and origin of nonbaryonic dark matter (DM).
Many theoretical models have predicted 
the Weakly Interacting Massive Particles (WIMP) 
of mass of the electroweak scale as DM candidates.
The WIMPs are assumed to be produced 
by the thermal freeze-out mechanism in the early Universe
and can be fermions or scalars or vector bosons
depending on the model.
As no signals of the WIMP have been observed 
in the high energy colliders 
and direct detection experiments, however,
much interest is devoted to the possibility
of light dark matter (LDM) in the keV--MeV mass range.
\cite{LDM}.

In this paper, we consider a fermionic LDM model 
which is suggested in Ref \cite{darkZ}
where the hidden sector is mediated by 
an additional SU(2) scalar doublet.
When we consider singlet fermions as DM candidates,
usually singlet scalars are introduced together
as a mediator between hidden sector and the SM
and an additional mass scale also introduced
by the vacuum expectation value (VEV) of the singlet scalar
\cite{SFDM1,SFDM2,SFDM3,SFDM4,SFDM5}.
However, we take the scalar doublet as a mediator,
neither singlet scalar nor new mass scale is required. 
Instead the U(1)$_X$ symmetry is required
for the fermionic DM candidate 
to couple to the mediator scalar doublet.
Thus the hidden sector in this model is QED-like,
which consists of a SM singlet fermion and 
a hidden U(1)$_X$ gauge field.
Since the mediator scalar is the SU(2) doublet 
and also carries the hidden U(1)$_X$ charge,
the U(1)$_X$ symmetry is broken by
the electroweak symmetry breaking (EWSB)
and the corresponding gauge boson gets its own mass
by the electroweak VEV.
The new massive gauge boson is mixed with the $Z$ boson
and is called the dark $Z$ boson.
This model satisfies strong electroweak constraints
from the low energy experiments 
and high energy collider phenomenology
on neutral current (NC) interactions.

It turns out that this model favors 
rather light dark $Z$ boson and fermionic DM.
If the DM candidate is lighter than the electron,
the o-Ps can annihilate into the DM pair 
through the dark $Z$ boson
and the final state is invisible in this model.
The dark $Z$ boson being light,
predictions of invisible decay rates of o-Ps into the DM pair 
can be much enhanced compared to 
the weak invisible decay rate in the SM,
which is a clear signal of the new physics.

On the other hand,
LDM with the mass less than MeV suffers from tension 
with cosmological observables
when it is in thermal equilibrium
with the bath of the SM particles in the early Universe
\cite{Boehm:2013jpa,Green:2017ybv,Sabti:2019mhn}.
It is because the temperature where the BBN started is affected
by extra relativistic degrees of freedom
and the predictions of the abundance of light elements
would be altered.
Recently Berlin and Blinov reported that 
sub-MeV LDM is allowed
when the equilibrium of the light state with the SM
is later than the neutrino decoupling
\cite{Berlin:2017ftj,Berlin:2018ztp}.
We take this scenario to accept our fermionic LDM candidate
lighter than the electron here.

In this paper we investigate
the exotic decays of Ps 
including invisible decays and single photon decays 
when the DM candidate is lighter than the electron 
in the singlet fermionic DM model 
with hidden U(1)$_X$ gauge group and 
an additional scalar doublet mediator.
The outline of this paper is as follows:
We briefly describe the model with
electroweak constraints in section 2.
In section 3, we present the predictions of the positronium decays 
in this model.
The dark matter phenomenology is elaborated 
in relation to the positronium decays in section 4.
We finally conclude in section 5.

\section{ Dark $Z$ Phenomenology}

The hidden sector of the model consists of
a SM gauge singlet Dirac fermion $\psi_X$ as a DM candidate
and a gauge field for a new U(1)$_X$ gauge symmetry.
We assume no kinetic mixing 
between the hidden U(1)$_X$ and the SM U(1)$_Y$
and the gauge charge of $\psi_X$ 
to be $(1,1,0,X)$ based on
${\rm SU}(3)_c \times {\rm SU}(2)_L \times 
{\rm U}(1)_Y \times {\rm U}(1)_X$ gauge group.
The SM fields do not carry the U(1)$_X$ gauge charge
and do not couple to the hidden sector fermion directly.

We introduce an additional SU(2) scalar doublet $H_1$ 
as a mediator field between the hidden sector and the SM sector
and the content of Higgs fields is the 2 Higgs Doublet Model (2HDM).
There are three free parameters,
U(1)$_X$ gauge coupling $g_X$
and the U(1)$_X$ charges of $H_1$ and $\psi_X$,
but they just appear in the form of $g_X X$.
Thus we have freedom to fix only one of three parameters.
We take the U(1)$_X$ charge of $H_1$ to be 1/2 
for convenience.  
Then we let the charge assignments of $H_1$ and 
the SM-like Higgs doublet $H_2$ 
be $ (1, 2, \frac{1}{2}, \frac{1}{2})$ for $H_1$,
and $ (1, 2, \frac{1}{2}, 0)$ for $H_2$, respectively.
Due to the U(1)$_X$ charge,
$H_1$ does not couple to the SM fermions
and the $H_2$ couplings to the SM fermions are same as
the SM Yukawa interactions
as in the Higgs sector in the 2HDM of type I. 
The Higgs sector lagrangian is given by
\be
{\cal L}_H = (D^\mu H_1)^\dagger D_\mu H_1 
	   + (D^\mu H_2)^\dagger D_\mu H_2 - V(H_1, H_2) 
	   + {\cal L}_{\rm Y}(H_2), 
\ee
where $V(H_1, H_2)$ is the Higgs potential,
${\cal L}_{\rm Y}$ the Yukawa couplings,
and the covariant derivative defined by
\be
D^\mu = \partial^\mu + i g W^{\mu a} T^a 
                 + i g' B^\mu Y + i g_X A_X^\mu X
\ee
with $T^a$ ($a=1,2,3$) being the SU(2) generators.
Here $X$ is the hidden U(1)$_X$ charge operator and 
$A_X^\mu$ the corresponding gauge field.
The Higgs potential is given by
\be
V(H_1,H_2) &=& \mu_1^2 H_1^\dagger H_1 + \mu_2^2 H_2^\dagger H_2
              + \lambda_1 (H_1^\dagger H_1)^2
	      + \lambda_2 (H_2^\dagger H_2)^2
\nonumber \\
	   && + \lambda_3 (H_1^\dagger H_1)(H_2^\dagger H_2)
              + \lambda_4 (H_1^\dagger H_2)(H_2^\dagger H_1),
\ee
where $\mu_{1,2}^2$ are dimension 2 couplings for
quadratic terms while $\lambda_{1,2,3,4}$
dimensionless quartic couplings.
Note that the soft $Z_2$ symmetry breaking terms of
the $H_1^\dagger H_2$ quadratic term 
and the quartic term $(H_1^\dagger H_2)^2$ with $\lambda_5$ coupling 
are forbidden by the U(1)$_X$ gauge symmetry.

The physical gauge bosons after the EWSB,
photon, $Z$ boson and the extra $Z$ boson ($Z'$) are defined by
\be
	A_X &=&
              c_X Z' + s_X Z, 
\nonumber \\
	W_3 &=&
             -s_X c_W Z' + c_X c_W Z + s_W A ,
\nonumber \\
	B &=&
	      s_X s_W Z' - c_X s_W Z + c_W A ,
\ee
where $s_W=\sin \theta_W=g'/\sqrt{g^2+{g'}^2}$, 
$c_W=\cos \theta_W$ 
is the Weinberg angle,
and $s_X=\sin \theta_X$, $c_X=\cos \theta_X$ 
the $Z-Z'$ mixing defined by
\beq
\tan 2 \theta_X 
                = \frac{-2 g_X g' s_W \cos^2 \beta }
		       {{g'}^2 - g_X^2 s_W^2 \cos^2 \beta}.
\eeq
The VEVs of two Higgs doublets,
$\langle H_i \rangle = (0,v_i/\sqrt{2})^T$ with $i=1,2$,
define $\tan \beta =v_2/v_1$.
We have the neutral gauge boson masses 
\be
m_{Z,Z'}^2 = \frac{1}{8} \left( g_X^2 v_1^2 +(g^2+{g'}^2) v^2
	       \pm \sqrt{(g_X^2 v_1^2 - (g^2+{g'}^2) v^2)^2
			 + 4 g_X^2 (g^2+{g'}^2) v_1^4} \right),
\ee
where $v^2=v_1^2+v_2^2$. 
Note that only two mixing angles are required to diagonalize
the neutral gauge boson mass matrix in this model.  

The NC interactions of $Z$ and $Z'$ bosons are given by
\be
{\cal L}_{NC} = 
	 \left( c_X {Z}^\mu + s_X {Z'}^\mu \right) 
	 \left( g_V \bar{f} \gamma_\mu f
	  + g_A \bar{f} \gamma_\mu \gamma^5 f \right),
\ee
where $g_V$ and $g_A$ are the SM $Z$ couplings to the fermions.
Since the $Z'$ interactions are same as the SM $Z$ interactions
except for the overall suppression by $s_X$,
we call it the dark $Z$ boson.

The $Z$ boson mass is shifted in this model such that
\be
m_Z^2 &=& \frac{m_W^2}{c_W^2 c_X^2} - m_{Z'}^2 \frac{s_X^2}{c_X^2},
\ee
which leads to the shift of the $\rho$ parameter
\be
\frac{1}{\rho} = \frac{m_Z^2 c_W^2}{m_W^2}
      = \frac{1}{c_X^2}
         - \frac{m_{Z'}^2 c_W^2}{m_W^2} \frac{s_X^2}{c_X^2}
      \approx 1 + s_X^2 \left(1 - \frac{m_{Z'}^2 c_W^2}{m_W^2} \right)      
      \equiv 1+\Delta \rho_{Z'}
\ee
in the leading order of $s_X^2$.
Moreover there exist also new scalar contributions
to the $\Delta \rho$ in this model,
given by 
\be
\Delta \rho^{(1)}_{\rm NS} = \frac{\alpha}{16 \pi m_W^2 s_W^2}
     \left( m_\pm^2 - \frac{m_H^2 m_\pm^2}{m_H^2-m_\pm^2}
              \log \left( \frac{m_H^2}{m_\pm^2} \right)
            \right)
\ee
at one-loop level \cite{hollik}. Here $m_H$ is the SM Higgs mass and 
$m_\pm$ is the charged Higgs mass. 
Then $\Delta \rho$ predicted in this model is
 $\Delta \rho_{\rm New} = \Delta \rho_{Z'} + \Delta \rho^{(1)}_{\rm NS}$.
The present limit on $\Delta \rho$ reads from
the measurements $ \alpha(m_Z)^{-1} = 127.955 \pm 0.010$
and Peskin-Takeuchi $T$ value
$T = 0.07 \pm 0.12$ \cite{PDG} as
\be
-0.00039 < \Delta \rho < 0.001485.
\ee
When $m_\pm \ge 120$ GeV,
$\Delta \rho_{\rm NS}$ exceeds 0.001485 and
no points of $(m_{Z'}, -s_X)$ are allowed.
Consequently 120 GeV is an upper limit on $m_\pm$ in this model.
Actually $\Delta \rho$ is insensitive to $m_\pm$ below 120 GeV.

\begin{figure}[t]
\centering
\includegraphics[width=12cm]{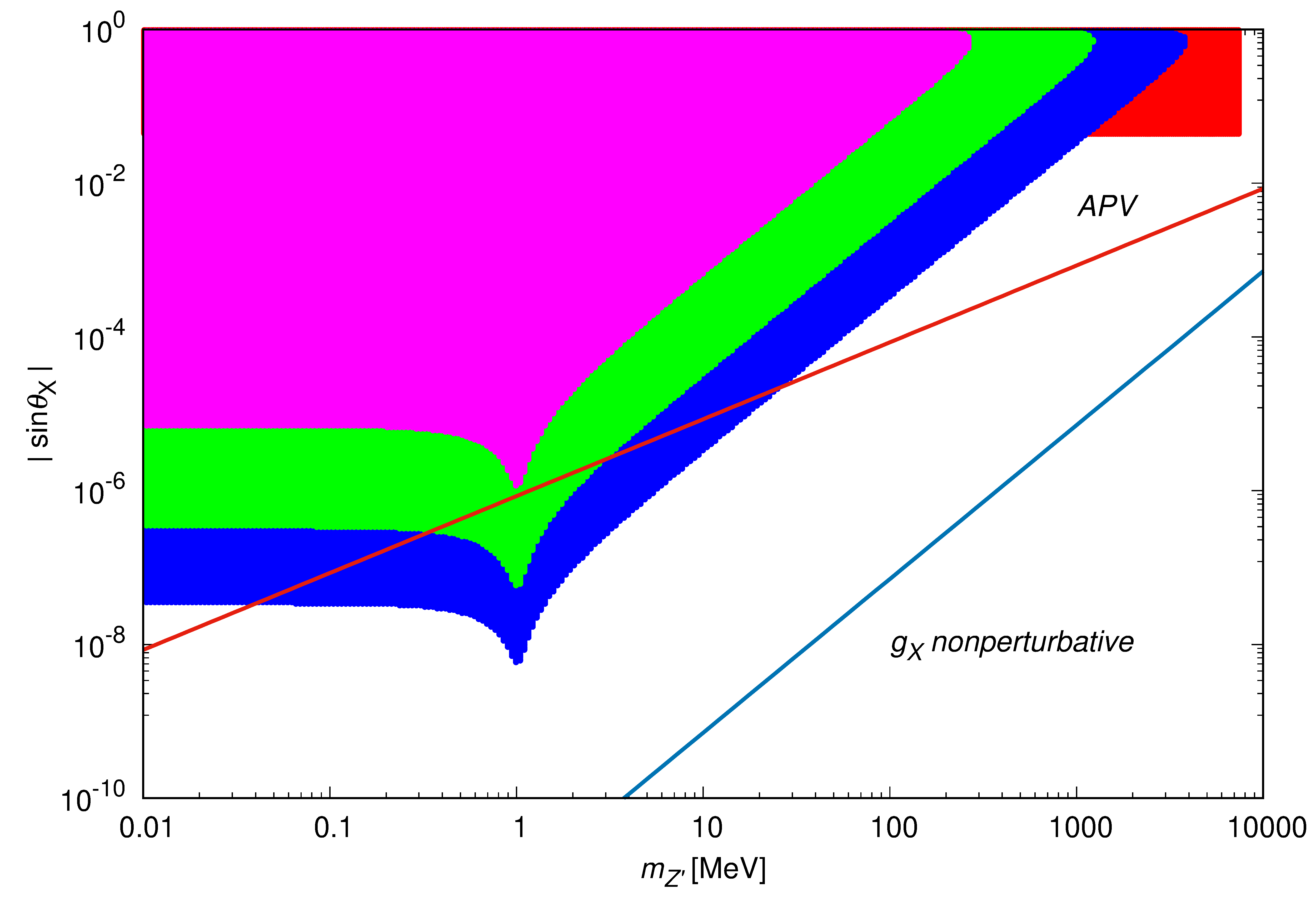}
\caption{
Exclusion regions by the positronium invisible decays
in the plane of the model parameter $(m_{Z'},|\sin \theta_X|)$.
The magenta region is excluded by the current limit 
Br(o-Ps $\to$ invisible)$<4.2 \times 10^{-7}$,
the green region by the future limit $< 10^{-9}$,
and the blue region by the future limit $< 10^{-11}$.
Electroweak constraints of $\Delta \rho$ is the red region 
(overlapped by the Ps exclusion regions)
and the APV constraints above the red line.
\label{fig:EWconstraints}
}
\end{figure}

The precise measurement of the atomic parity violation (APV)
also provides a strong constraint on the exotic NC interactions.
The dark $Z$ exchange also contributes to
the shift of the weak charge
\beq
Q_W = Q_W^{SM} \left( 1+\frac{m_Z^2}{m_{Z'}^2} s_X^2 \right)
\eeq
in the leading order of $s_X$.
The SM prediction of the Cs atom is
$ Q_W^{SM} =  -73.16\pm0.05$ \cite{APVSM1,APVSM2},
and the present experimental value is
$ Q_W^{exp} = -73.16\pm0.35$ \cite{APV},
which yields the bound
\beq
\frac{m_Z^2}{m_{Z'}^2} s_X^2 \le 0.006
\eeq
at 90 \% C.L..
The $\Delta \rho $ and APV constraints are depicted 
in Fig.~\ref{fig:EWconstraints} 
by the red region and red line, respectively.

We assume that the hidden sector lagrangian is QED-like,
\be
{\cal L}_{\rm hs} = -\frac{1}{4} F_X^{\mu \nu} F_{X \mu \nu} 
                    + \bar{\psi}_X i \gamma^\mu D_\mu \psi_X
                          - m_X \bar{\psi}_X \psi_X,
\ee
where
\be
D^\mu = \partial^\mu + i g_X A^\mu_X X.
\ee
After the EWSB, the U(1)$_X$ is broken
and $\psi_X$ is connected to the SM through $Z$ and dark $Z$ bosons,
given in Eq.~(5).
Note that the fermion mass $m_X$ is a free parameter.
We will set $m_X < m_e$ in the next section.

\section{Positronium Decays}

In this model Ps can annihilate into 
the DM fermion pair through the dark $Z$ boson
when the DM fermion is lighter than the electron, $m_X < m_e$.
We obtain the dark $Z$ contribution to the invisible decays as
\be
\Gamma({\rm o\textrm{-} Ps} \to Z' \to \bar{\psi}_X \psi_X) 
       &=& \frac{1}{12 \pi m_e^2}
		    s_X^2 c_X^2 g_V^2 (g_X X)^2 
	   \left[ \left( 1-\frac{m_{Z'}^2}{4 m_e^2} \right)^{2}
		    + \frac{m_{Z'}^2 \Gamma_{Z'}^2}
		           {16 m_e^4}
	   \right]^{-2}
\nonumber \\
            &&  ~~~~~~~~~~\times \sqrt{1-\frac{m_X^2}{m_e^2} }
		    \left( 1+\frac{m_X^2}{2 m_e^2} \right)
          |\psi(0)|^2,
\ee
where the square of the Ps wavefunction at the origin is
$|\psi(0)|^2=m_e^3 \alpha^3/8 \pi$.
The dark $Z$ couplings to the SM particles are
suppressed by the mixing angle $s_X$
while not suppressed to the DM fermions.
Thus
the decay width of the dark $Z$ boson is dominated by
$Z' \to  \bar{\psi}_X \psi_X$, 
where $\Gamma_{Z'}=\Gamma(Z' \to \bar{\psi}_X \psi_X) \sim 10^{-2}$ MeV, 
when $m_{Z'} > 2 m_X$.
On the other hand, if $m_{Z'} < 2 m_X$, then
only the neutrino channels are allowed and
$ \Gamma_{Z'} = 3 \Gamma(Z' \to \nu \bar{\nu}) 
= 3 \Gamma(Z \to \nu \bar{\nu})(s_X/c_X)^2 (m_{Z'}/m_Z) < 10^{-13}$ MeV 
at most. 
Therefore we can neglect $\Gamma_{Z'}$ to calculate
the invisible decay rate except for around the resonance region.

The branching ratio of the o-Ps invisible decay is 
\be
{\rm Br}({\rm o\textrm{-} Ps} \to {\rm invisible}) 
= \frac{\Gamma({\rm o\textrm{-} Ps} 
\to {\rm invisible})}{\Gamma_0 + \Gamma({\rm o\textrm{-} Ps} 
\to {\rm invisible})},
\ee
where the SM decay rate $\Gamma_0$ is dominated by
the three photon decay given by
\be
\Gamma({\rm o\textrm{-} Ps} \to \gamma \gamma \gamma)
     = \frac{2(\pi^2-9) m_e \alpha^6}{9 \pi}
     \left(1-10.28661 \frac{\alpha}{\pi} + {\cal O}(\alpha^2) \right)
     \approx 7.0382 \,\mu {\rm s}^{-1}.
\ee
Figure~\ref{fig:EWconstraints} depicts the exclusive region by 
the invisible positronium decays 
with the present data of Eq.~(\ref{broPs})
and the future experimental reaches 
Br(o-Ps$\to$invisible)$< 10^{-9}$ and $< 10^{-11}$
\cite{KAPAE}.
This model has two more free parameters on the DM sector,
the hidden U(1)$_X$ charge and the mass of $\psi_X$. 
We take $(g_X X)^2=2 \pi$ and $m_X=m_e/2$ 
in Fig.~\ref{fig:EWconstraints}
as benchmark values
which are chosen to probe the parameter region 
below the APV bound and 
to accommodate the DM phenomenology 
as will be discussed in the next section.

The para-positronium dominantly
decays into two photons.
In this model, p-Ps can decay into a photon and a dark $Z$ boson, 
then we will observe the single photon decay process.
The decay rate into $\gamma Z'$ is
\be
\Gamma({\rm p\textrm{-} Ps} \to \gamma Z') 
       = \frac{2\alpha }{m_e^2} s_X^2 g_V^2 
            \left( 1 -\frac{m_{Z'}^2}{4m_e^2}\right)
          |\psi(0)|^2.
\ee
Meanwhile o-Ps can also decay into the $\gamma Z'$ final state 
due to the axial coupling of the dark $Z$ such as
\be
\Gamma({\rm o\textrm{-} Ps} \to \gamma Z')
       = \frac{8\alpha }{3 m_{Z'}^2} s_X^2 g_A^2 
		\left( 1 -\frac{m_{Z'}^2}{4m_e^2}\right)
		\left( 1 +\frac{m_{Z'}^2}{4m_e^2}\right)
          |\psi(0)|^2.
\ee
We estimate the branching ratios
Br(p-Ps$\to \gamma Z') < 10^{-12}$
and Br(o-Ps$\to \gamma Z') < 10^{-13}$
with the allowed values of $(m_{Z'}, |\sin \theta_X|)$
given in Fig.~\ref{fig:EWconstraints},
which are smaller than the future experimental reaches 
considered in Fig.~\ref{fig:EWconstraints}.
Neither the experimental limits for the single photon decay of p-Ps
nor those of o-Ps have not been reported yet.

\section{Dark Matter Phenomenology}

We calculate the relic abundance $\Omega_{\rm CDM} h^2$ 
in the thermal freeze-out scenario 
using the \texttt{micrOMEGAs} 
\cite{micromegas} with the allowed values 
of parameters $(m_{Z'}, |\sin \theta_X|)$
given in the previous section
and show that the model prediction can accommodate
the present measurements with high precision
\cite{PDG}
\beq
\Omega_{\rm CDM} h^2 = 0.1186 \pm 0.0020
\eeq
from the anisotropy of the cosmic microwave background (CMB) 
and of the spatial distribution of galaxies.
Since we are interested in the parameter set 
with which the dark sector has a sizable effect 
on the positronium physics, 
we consider a light DM (LDM) scenario 
that constrains the DM mass smaller than the electron mass. 
In this case, the model is barely constrained by both
the direct \cite{Xenon1t, CRESST, Darkside, LUX} 
and indirect \cite{FermiLAT, HESS} detection experiments 
as discussed in Ref. \cite{darkZ}.

\begin{figure}[t]
\centering
\includegraphics[width=10cm]{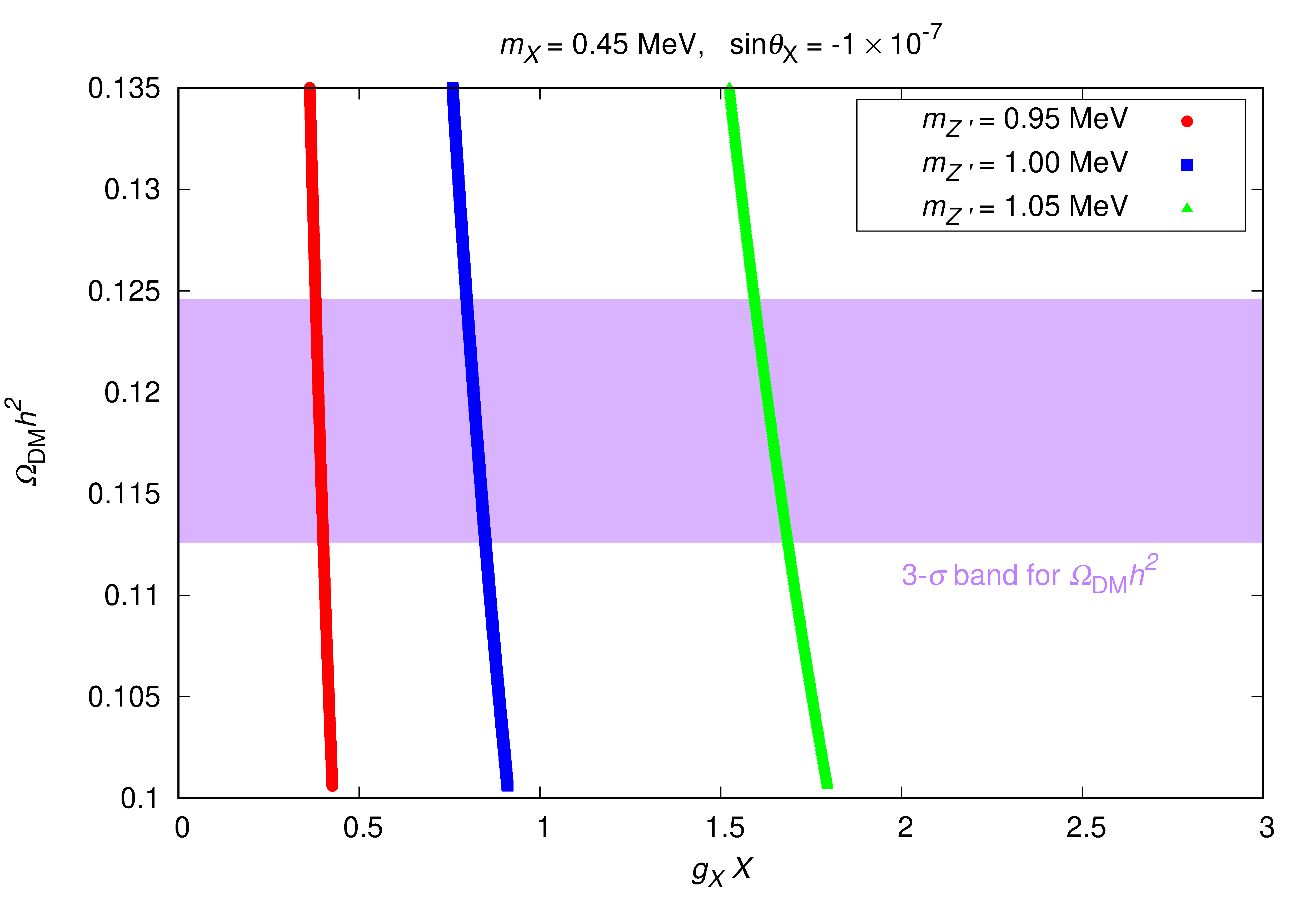}
\caption{
Behaviors of relic density for a few benchmark points, 
as functions of the interaction strength 
between DM pairs and $Z'$ boson.
}
\label{fig:benchmark}
\end{figure}

\begin{figure}[h]
\centering
\includegraphics[width=10cm]{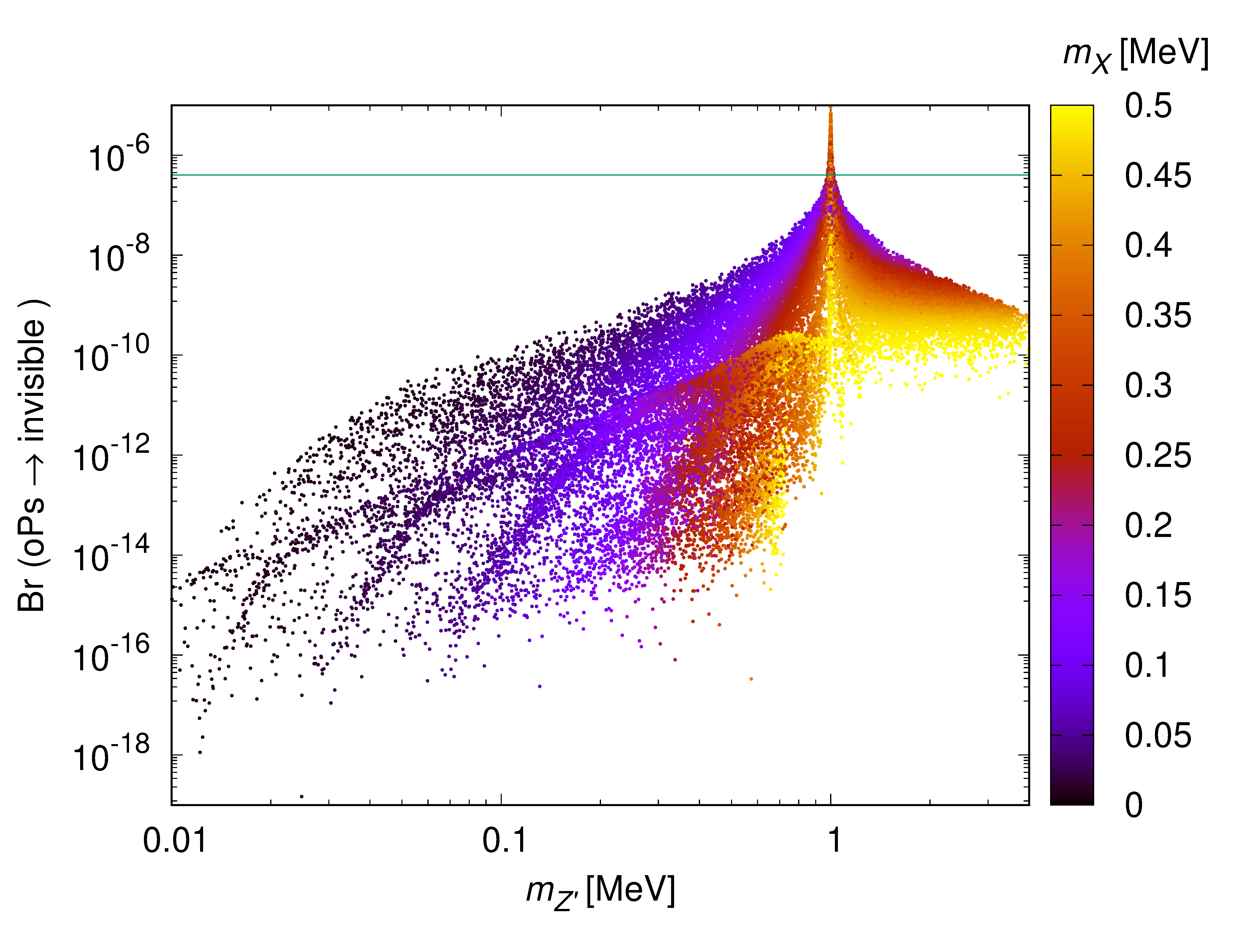}
\caption{
Branching ratios of the o-Ps invisible decay for points 
that accommodate the observed relic density of DM 
within $3\sigma$ range as a function of $Z'$ mass 
in unit of MeV.
In the plot, two distinct kinds of points are overlapped. 
One is the case of $m_{Z'} < 2 m_X$, 
where the decay width of $Z'$ is relatively suppressed 
compared with the opposite case, $m_{Z'} > 2 m_X$. 
Green horizontal line corresponds to the present limit 
on the branching ratio of the o-Ps invisible decay, 
$4.2 \times 10^{-7}$.
}
\label{fig:invisible}
\end{figure}

We demonstrate generic behaviors of $\Omega_{\rm CDM} h^2$ 
for some benchmark points near the resonance region of Ps decay
in Fig.~\ref{fig:benchmark}.
We find that the relic abundance is very sensitive to the value of 
$g_X X$ in the resonance region around $m_{Z'}=1$ MeV.
In Fig.~\ref{fig:invisible},
we show the branching ratios of the o-Ps invisible decay 
for points that accommodate the present relic density 
with $3\sigma$ range as a function of the 
$Z'$ mass in unit of MeV.
For this analysis, we scan the DM mass less than $0.5$ MeV 
and the $Z'$ mass in the range $0.01-4$ MeV 
to investigate possible correlations 
between the DM and positronium physics. 
As shown in Fig.~\ref{fig:invisible}, 
the branching ratio of o-Ps invisible decay 
can be enhanced significantly in the resonance region 
around $m_{Z'} \simeq 1$ MeV.
Though relic density depends on various model parameters, 
it is largely determined by the interaction strength 
between DM pair and $Z'$ boson, $g_X X$, generically.
Perturbativity bound on the size of the coupling 
between the DM pair and $Z'$ is also imposed.

\begin{figure}[t]
\centering
\includegraphics[width=7cm]{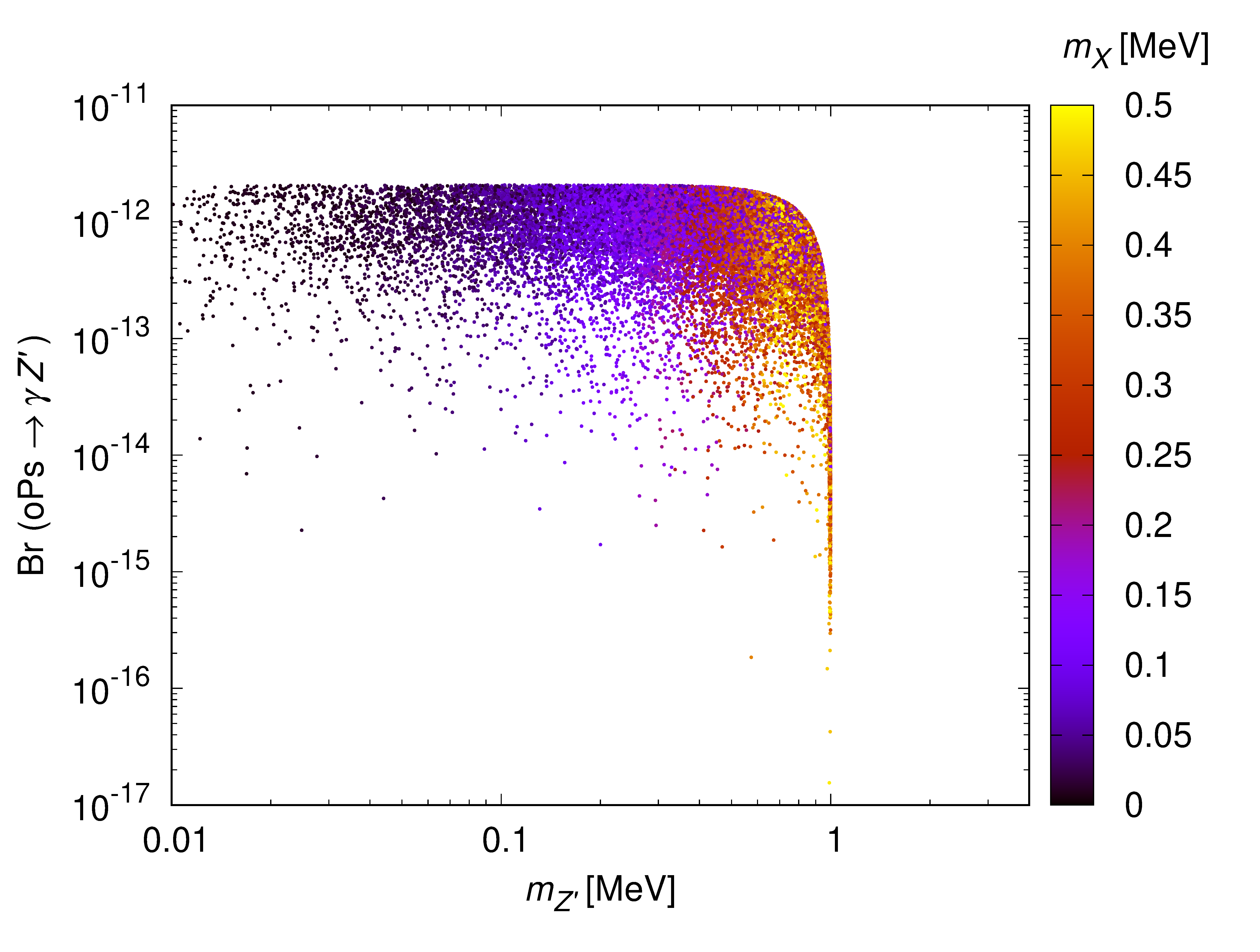}
\includegraphics[width=7cm]{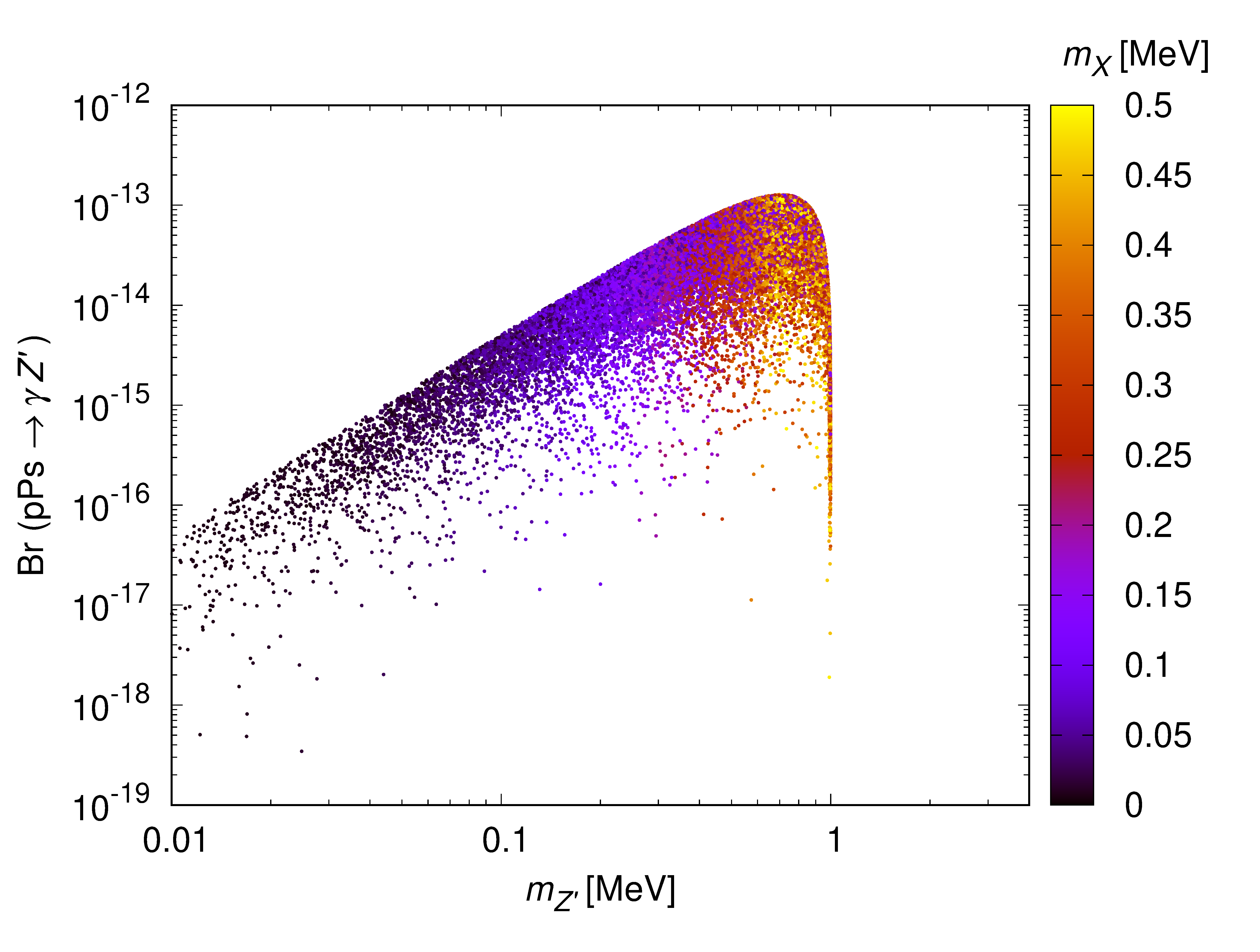}
\caption{
Branching ratios of o-Ps and p-Ps decay 
into a photon and a $Z'$ boson for points 
that accommodate the observed relic density of DM 
within $3\sigma$ range as a function of the $Z'$ mass in unit of MeV.
They appear as the conversion of positronium to single photon.
}
\label{fig:Zpgamma}
\end{figure}

We also compute the branching ratios of 
o-Ps and p-Ps decays into a photon and a $Z'$ boson, 
that will be observed as single photon decays, 
with parameter sets allowed 
by the electroweak constraints and DM relic density
and show results 
in Fig.~\ref{fig:Zpgamma}.
The predicted branching ratios of o-Ps and p-Ps are 
at most $O(10^{-12})$ and $O(10^{-13})$
as estimated in the previous section, 
which are rather far from the reach of precision 
for the search of the invisible decays of positronium 
in near future.

A few comments are in order. 
Generically, dark matter mass below $1$ MeV 
is disfavored by BBN and CMB data
through modifying the effective number of neutrino species 
when employing conventional thermal freeze-out scenarios 
\cite{Green:2017ybv,Serpico:2004nm,Ho:2012ug}. 
Recently the authors of  
Ref. \cite{Berlin:2017ftj,Berlin:2018ztp} 
suggested an alternative cosmological scenarios that 
can alleviate the problem for sub-MeV DM. 
It is dubbed as `delayed equilibration scenario' 
in which sub-MeV DM thermalizes with the SM sector 
below the neutrino-photon decoupling temperature. 
In this scenario, the SM bath is cooled down 
by the equilibration and is heated again by the freeze-out of DM.
In the dark $Z$ model, 
a DM pair can convert to an electron-positron pair 
through $Z'$ to be constrained from the supernova.
However after comprehensive analysis
about the supernova constraints 
on the dark photon portal models 
the authors in Ref.~\cite{Sung:2021swd} found that 
such constraints can be evaded 
for a dark sector with dark fine structure constant 
$\alpha_D \gtrsim 10^{-7}$. 
We find that the dark $Z$ model, in which DM coupling to $Z'$ 
is taken to be large enough to accommodate the present relic density 
as in Fig.~\ref{fig:benchmark}, 
can avoid the supernova constraints. 
We conclude that the parameter region studied in this paper 
is still valid. 

\section{Concluding Remarks}

We study the invisible decay of o-Ps into fermionic DM pair
through the dark $Z$ boson when the fermionic DM particle
is lighter than the electron.
Still the predictions of the Ps invisible decay rates
in the dark $Z$ model
are less than the present experimental reach
but much enhanced compared to the SM predictions
through the weak interaction.
The fermionic LDM scenario with the light mediator
can also satisfy the relic abundance
and is not constrained by the present
direct detection experiment and indirect observations.
We discuss that the LDM model discussed in this work
can be accommodated in the recent delayed equilibration scenario.
In conclusion the Ps invisible decay provides attractive
phenomenology of the dark $Z$ model with fermionic DM
which is independent of collider and dark matter phenomenology.

\acknowledgments
This work is supported 
by the Gyeongsang National University Fund 
for Professors on Sabbatical Leave, 2020 (KYL).
This work is also supported 
by Basic Science Research Program
through the National Research Foundation of Korea 
funded by the Ministry of Science and ICT
under the Grants 
No. NRF-2020R1A2C3009918 (DWJ),
No. NRF-2020R1I1A1A01073770 (CY), 
and also funded by the Ministry of Education
under the Grants No. NRF-2018R1D1A1B07047812 (DWJ).

\def\PRDD #1 #2 #3 {Phys. Rev. D \textbf{#1},\ #2 (#3)}
\def\PRD #1 #2 #3 #4 {Phys. Rev. D \textbf{#1},\ No. #2, #3 (#4)}
\def\PRLL #1 #2 #3 {Phys. Rev. Lett. {\bf#1},\ #2 (#3)}
\def\PRL #1 #2 #3 #4 {Phys. Rev. Lett. {\bf#1},\ No. #2, #3 (#4)}
\def\PRA #1 #2 #3 {Phys. Rev. A {\bf#1},\ #2 (#3)}
\def\PLB #1 #2 #3 {Phys. Lett. B {\bf#1},\ #2 (#3)}
\def\NPB #1 #2 #3 {Nucl. Phys. B {\bf #1},\ #2 (#3)}
\def\ZPC #1 #2 #3 {Z. Phys. C {\bf#1},\ #2 (#3)}
\def\EPJ #1 #2 #3 {Euro. Phys. J. C {\bf#1},\ #2 (#3)}
\def\JPG #1 #2 #3 {J. Phys. G: Nucl. Part. Phys. {\bf#1},\ #2 (#3)}
\def\JHEP #1 #2 #3 {JHEP {\bf#1},\ #2 (#3)}
\def\JCAP #1 #2 #3 {JCAP {\bf#1},\ #2 (#3)}
\def\IJMP #1 #2 #3 {Int. J. Mod. Phys. A {\bf#1},\ #2 (#3)}
\def\MPL #1 #2 #3 {Mod. Phys. Lett. A {\bf#1},\ #2 (#3)}
\def\PTP #1 #2 #3 {Prog. Theor. Phys. {\bf#1},\ #2 (#3)}
\def\PR #1 #2 #3 {Phys. Rep. {\bf#1},\ #2 (#3)}
\def\RMP #1 #2 #3 {Rev. Mod. Phys. {\bf#1},\ #2 (#3)}
\def\PRold #1 #2 #3 {Phys. Rev. {\bf#1},\ #2 (#3)}
\def\IBID #1 #2 #3 {{\it ibid.} {\bf#1},\ #2 (#3)}
\def\etal {{\it et al.} }

\end{document}